\documentclass[preprint,12pt,sort&compress]{elsarticle}

\usepackage{graphicx}
\begin{document}

\begin{frontmatter}



\title{Parton radiative processes and pressure isotropization 
in relativistic heavy ion collisions}


\author{Bin Zhang and Warner A. Wortman}

\address{Department of Chemistry and Physics, 
Arkansas State University,\\
P.O. Box 419, State University, AR 72467-0419, U.S.A.}

\begin{abstract}
The impact of radiative processes on kinetic 
equilibration is studied via a radiative transport 
model. The $2\leftrightarrow 3$ processes can
significantly increase the level of 
thermalization. These processes lead to 
an approximate coupling constant scaling of the 
evolution of the pressure anisotropy qualitatively 
different from the case with only $2\rightarrow 2$ 
partonic processes. Furthermore, thermal and 
Color Glass Condensate motivated initial conditions 
are shown to share the same asymptotic evolution when
$2\leftrightarrow 3$ processes are included.
This emphasizes the unique role of radiative
processes in Quark-Gluon Plasma thermalization.
\end{abstract}


\end{frontmatter}

% \linenumbers

\section{Introduction}
\label{sec:intro}

The Quark-Gluon Plasma can be produced in central
relativistic heavy ion collisions 
\cite{Adams:2005dq,Adcox:2004mh,Back:2004je,Arsene:2004fa}. 
It leads to 
strong collective flow and jet quenching. Ideal
hydrodynamics is very successful in describing
experimental observables in the low transverse
momentum region \cite{Teaney:2000cw,Huovinen:2001cy,Kolb:2001qz,Hirano:2005xf}. Transport models can also give
good descriptions of global observables in these
collisions \cite{Bass:2002fh,Molnar:2001ux,Zhang:1999bd,Lin:2000cx,Lin:2001yd,Lin:2004en,Xu:2004mz,Ferini:2008he}. Ideal hydrodynamics assumes local
thermal equilibrium and the ideal hydrodynamic equations are
valid when there is local isotropization \cite{Heinz:2009xj}. It is
interesting to study the effects of viscosity 
\cite{Muronga:2001zk,Dusling:2007gi,Luzum:2008cw,Song:2007ux,Song:2008si,Song:2007fn,Denicol:2010xn}
and how the partons produced
in heavy ion collisions thermalize \cite{Berges:2005ai,Florkowski:2008ag,Chesler:2008hg,Chesler:2009cy}. 
The equilibration
process can be studied with a microscopic transport
model \cite{Xu:2007aa,Zhang:2008zzk,Huovinen:2008te,El:2009vj}. 
In particular, Xu and Greiner studied the transport
rates \cite{Xu:2007aa} and showed that radiative
processes are important for parton momentum isotropization.
A related topic is pressure isotropization. 
The Frankfurt group extended the viscous hydrodynamic equation
to the third order \cite{El:2009vj} for
the pressure isotropization study. 
They compared pressure anisotropy evolutions
starting from an isotropic initial condition with both
the extended viscous hydrodynamics and 
an elastic parton cascade with time dependent cross 
sections. The comparison demonstrates the importance 
of higher order corrections in describing 
highly viscous matters.
In the following, we will study whether and how 
pressure isotropization depends on the 
inclusion of radiative processes. The pressure
anisotropy evolutions with different initial
anisotropies, energy densities, fugacities, momentum
distributions, coupling constants will be compared
and contrasted to demonstrate the effects of radiative
processes. We will further discuss how the system loses 
memory of the initial pressure anisotropy to approach the
common asymptotic evolution and the interplay
between chemical equilibration and kinetic equilibration.

\section{Pressure Isotropization and radiative transport}
\label{sec:thema}

Relativistic heavy ion collisions produce hot and dense
matter. The highest energy density is achieved in the central
cell in central collisions. Kinetic equilibration in the 
central cell can be 
characterized by the pressure anisotropy parameter, i.e.,
the longitude pressure to the transverse pressure 
ratio $P_L/P_T$.
When there is thermal equilibrium, the pressure anisotropy
equals one. Any pressure anisotropy value that is not 
equal to one indicates non-equilibrium conditions.

We will study the early stage when longitudinal expansion
dominates the central cell evolution. 
The initial conditions will be taken to be similar to
those expected in central relativistic heavy ion collisions.
Only massless gluons are included and the formation
proper time is set to be 0.5 fm/$c$. These gluons
will be distributed uniformly
inside a transverse circle with a radius of 5 fm and a
space-time rapidity region from -5 to +5. The initial
space-time rapidity density $dN/d\eta_s=1000$. 
The partonic system can start with an initial
energy density $\epsilon_0=38.2$ GeV/fm$^3$
or $76.4$ GeV/fm$^3$. In the local rest frame, the
initial particle distribution can be isotropic or 
transverse. Both have exponential momentum distributions.
In the isotropic case,
$\epsilon_0=38.2$ GeV/fm$^3$ is equivalent
to an initial temperature of $T_0=0.5$ GeV and 
$\epsilon_0=76.4$ GeV/fm$^3$ corresponds to an initial
temperature of $1$ GeV. 
The transverse initial conditions are isotropic 
in the transverse plane with an
initial temperature 50\% higher than the isotropic
case to have the same initial energy density.

To study the effects of radiative processes, 
we will compare results with
only the $2\rightarrow 2$ process
that has 2 incoming gluons and 2 outgoing gluons
with those including
additional $2\leftrightarrow 3$ processes. The
$2\rightarrow 2$ cross section is set to be
the perturbative QCD
cross section regulated by a Debye screening mass, i.e.,
\begin{equation}
\sigma_{22}=\frac{9\pi\alpha_s^2}{2\mu^2}.
\end{equation}
In the above formula, $\alpha_s$ is the strong interaction
coupling constant and $\mu$ is the Debye screening mass.
The Debye screening mass is calculated dynamically via
\begin{equation}
\mu^2=\frac{3\pi\alpha_s}{V}\sum_i\frac{1}{p_i},
\end{equation}
where $V$ is the volume of the cell and the sum goes over
all gluons in the cell.

%dynamical screening and thermalization
The effect of dynamical screening can be demonstrated
by comparing the collision rate with the expansion
rate. In a fixed box, evolution from a pressure
anisotropy different from $1$ is characterized by
the collision rate only. However, when the system
undergoes longitudinal expansion, there is a 
tendency of evolving toward $0$ pressure anisotropy.
The initial
pressure anisotropy and the expansion rate
are responsible for how fast this
happens. It is the competition between collision
and expansion that determines whether pressure
isotropization can happen.
For the simple case when the system can be approximately
described by a temperature $T$, the cross section
$\sigma\propto 1/\mu^2 \propto 1/T^2$. The collision
rate $R_c=n\sigma\propto T^3\times1/T^2
\sim T\propto 1/\tau^{1/3}$. In the above
formula, $n$ stands for the particle number density
and the relation between $T$ and the proper time
$\tau$ for the Bjorken expansion is used. The volume
expansion rate $R_v\propto 1/\tau$. It decreases
faster than the collision rate. Therefore,
even if the initial expansion rate is 
large and there is a large initial
pressure anisotropy, resulting in
a decrease of pressure anisotropy
toward $0$, the collision rate will eventually take
over and pressure anisotropy will evolve toward $1$.

The $2\leftrightarrow 3$ radiative processes are implemented
for the study of the importance of radiative processes.
In particular, the $2\rightarrow 3$ cross section is taken to
be $1/2$ of the $2\rightarrow 2$ cross section. This is 
in line with a more elaborate study by 
Xu and Greiner \cite{Xu:2004mz}. To ensure
the correct chemical equilibrium, the detailed balance
relation needs to be enforced for the $3\rightarrow 2$ process.
We will take all outgoing particles to be isotropic in the
center of mass frame of the collision. This
is expected to reflect the gross features of particle
production in the central region. The $3\rightarrow 2$ collision
rate is related to the reaction integral $I_{32}$, i.e.,
the integral 
over the phase space volume (with proper summation and averaging
over internal degrees of freedom) 
of the modulus squared of the transition matrix element $M$.
In this case,
$I_{32}=\frac{1}{2!}\frac{d^2}{(2\pi)^{3\times2}}
|M|^2(2\pi)^4R_2(s^{1/2},0,0)$, where $d=16$ is the gluon
degeneracy factor, $R_2$ is the two-body phase space 
integral as defined in \cite{Hagedorn:1964rk}, 
and $s$ is the center of mass
energy squared. It is directly proportional 
to the $2\rightarrow 3$ cross section, i.e.,
\begin{equation}
I_{32}=\frac{192\pi^2}{d}\sigma_{23}=12\pi^2\sigma_{23}.
\end{equation}

%particle production and reaction rate per particle
The inclusion of radiative processes couples the chemical
equilibration to the kinetic equilibration. If the system
is not far from chemical equilibrium, then the inclusion
of radiative processes will lead to additional collisions
and isotropization. If the system is far below chemical
equilibrium, there will be significant particle
production. This will lead to smaller cross sections
which will limit the additional isotropization relative
to the case with only elastic processes. If the system is
far above chemical equilibrium, the decrease in particle
number will lead to larger cross sections and enhanced
isotropization.

\begin{figure}
\centering
\includegraphics{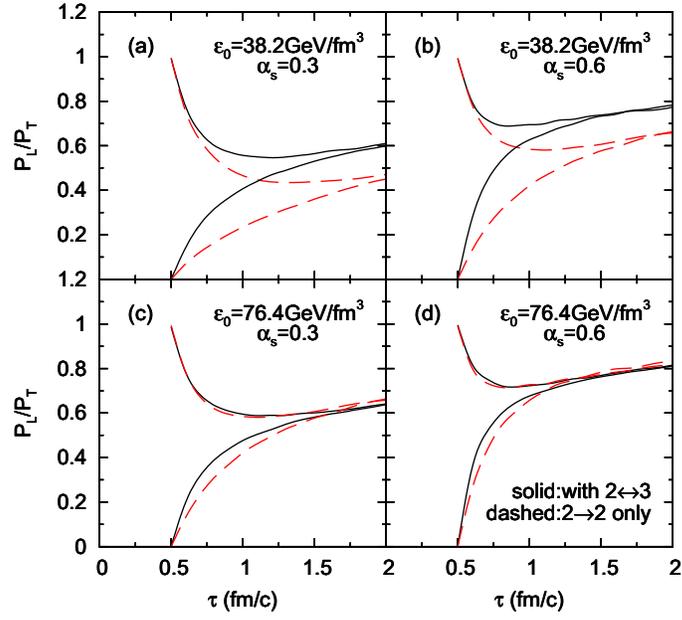}
\caption{Pressure anisotropy evolution. Solid curves
are for the case including the $2\leftrightarrow 3$ processes
while dashed curves are for the case with $2\rightarrow 2$
only. The upper panels have initial energy density 
$\epsilon_0=38.2$ GeV/fm$^3$ while the lower panels have 
$\epsilon_0=76.4$ GeV/fm$^3$. The left panels are calculated
with $\alpha_s=0.3$ while the right panels have $\alpha_s=0.6$.
}
\label{fig:plopt1}
\end{figure}

Pressure anisotropy evolutions with elastic collisions only
and with radiative processes are compared in 
Fig.~\ref{fig:plopt1}. First, let us focus on panel (b) and
discuss some of the general features of the evolutions. 
There are two curves for the case including 
$2\leftrightarrow 3$. One is for the case starting with
an initial isotropic distribution, i.e., initial
$P_L/P_T=1$. The other is for the case with an inside-outside
type of initial distribution, i.e., initial $P_L/P_T=0$.
Even though they start from quite different initial
pressure anisotropies, they approach the same asymptotic
evolution at late times. In other words, the memory
of initial anisotropy is lost after some characteristic time.
The case
with only $2\rightarrow 2$ collisions has approximately
the same behavior. The two curves approach and meet
each other at late times. This feature is the same
for all other cases. The importance of radiative processes
is reflected in the difference of the radiative case
and the case with only elastic collisions. 
The radiative case
has a pressure anisotropy of about 0.78 at 2 fm/$c$,
significantly larger than that of the elastic only case
of 0.65. This shows that radiative processes can
significantly enhance thermalization.

When $\alpha_s$ decreases from $0.6$ to $0.3$, the 
asymptotic value at 2 fm/$c$ decreases for both
the case with $2\leftrightarrow 3$ and without
$2\leftrightarrow 3$. When the initial energy density
increases from $38.2$ GeV/fm$^3$, the case with
$2\leftrightarrow 3$ increases slightly leading to
an approximate $\alpha_s$ scaling insensitive to
the initial energy density. 
In contrast, the case with only 
the $2\rightarrow 2$ process
increases drastically. 
Because of the slight increase in the radiative case
and drastic increase in the elastic only case as
the energy density increases, when the initial
energy density $\epsilon_0=76.4$ GeV/fm$^3$, the
radiative and elastic only cases have about the
same pressure anisotropy at $2$ fm/$c$. 
The elastic only case is sensitive to both the 
initial energy density and the coupling constant.
The large initial energy density
and small coupling constant case is related to
the small initial energy density and large coupling
constant case via the $\epsilon_0 \alpha_s$
scaling. The elastic only
curves in panel (b) and panel (c) follow this
$\epsilon_0 \alpha_s$ scaling and are identical
for these two cases.

The above analysis demonstrates that evolutions 
with and without radiative
processes have different dependences on system
parameters. In the case with radiative processes,
when the energy density is increased, the 
screening mass decreases, increasing the
cross section and collision rate. However,
this is counteracted by the additional particle
production. As the system falls far below 
chemical equilibrium, additional particle
production tends to increase the screening
mass and reduce the collision rate.
Lacking this balancing factor, the elastic only
case cannot maintain approximately the same
evolution as with the lower energy density
case and the pressure anisotropy increases
drastically. Decreasing the coupling constant
can compensate for the increase in pressure
anisotropy caused by the energy density increase.
As the initial binary collision cross
section is proportional to 
$\epsilon_0 \alpha_s$ with fixed initial number
density, when the coupling 
constant is decreased by the same factor
as the energy density increase, there
is an exact $\epsilon_0 \alpha_s$ scaling in
the elastic only case.
This is qualitatively different from the
approximate $\alpha_s$ scaling in the radiative case. 

\begin{figure}
\centering
\includegraphics{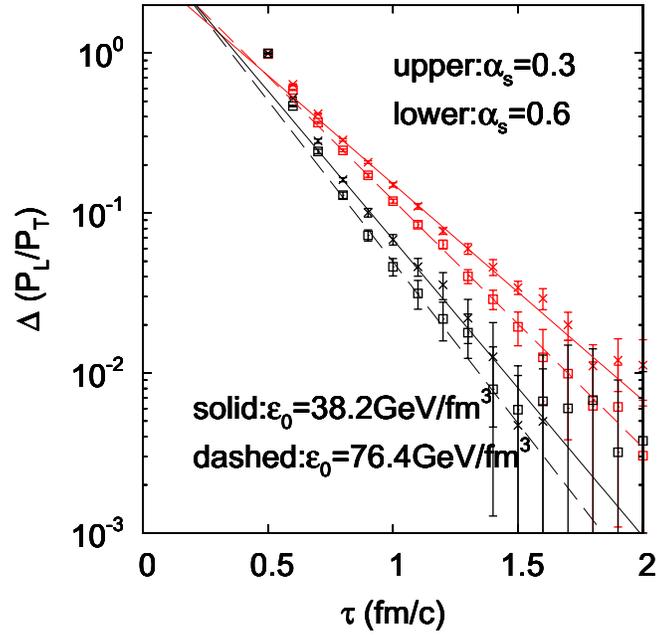}
\caption{Pressure anisotropy difference evolution. 
These results all include the radiative processes. Crosses 
(squares) are for initial energy density 
$\epsilon_0=38.2 (76.4)$ GeV/fm$^3$. Solid and dashed 
curves are exponential fits for $\tau \ge 0.8$ fm/$c$. 
For the two data sets with the same symbol, the upper
one is for $\alpha_s=0.3$ and the lower one is for 
$\alpha_s=0.6$.
}
\label{fig:dplopt}
\end{figure}

The two data sets with different initial pressure anisotropies
but otherwise same parameters seem to converge toward a common
final evolution. The difference between these two data sets
can be used to study the rate of convergence as shown in 
Fig.~\ref{fig:dplopt} for cases including the 
$2\leftrightarrow 3$ processes. The difference quickly
decreases and at late times fits well with an exponential 
decrease with proper time. In the case with $2\rightarrow 2$
only, the two curves may cross each other at late
times (see e.g. panel (d)
in Fig.~\ref{fig:plopt1}). In any case, by 
2 fm/$c$, the differences all go to about or below 
1\% of the initial 
difference. The system approaches the asymptotic evolution
on a very short time scale. The comparison in 
Fig.~\ref{fig:dplopt} shows that the approach to 
asymptotic evolution is faster with higher initial
energy density or higher coupling constant. It is interesting
to notice that within error bars, these lines seem to go
through the same point. We also looked at the curves 
with only elastic processes. They
also seem to go through a common point, even though
that point is different from the one with inelastic processes.

\begin{figure}
\centering
\includegraphics{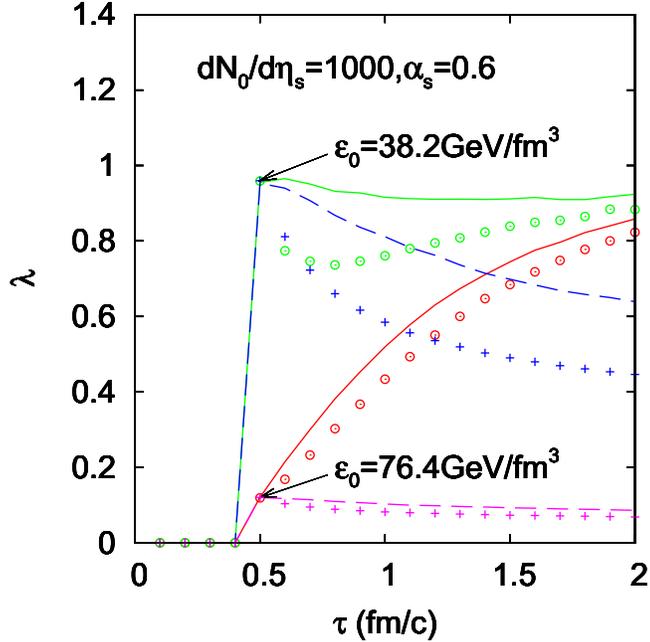}
\caption{Proper time evolution of fugacity $\lambda$ for
initial space-time rapidity density $dN_0/d\eta_s=1000$
and coupling constant $\alpha_s=0.6$. 
Lines are for isotropic exponential initial conditions
and points are for transverse exponential initial conditions.
Solid lines and circles are results including 
$2\leftrightarrow 3$. Dashed lines and pluses are results
with $2\rightarrow 2$ only. For the two data sets with
the same symbol, the upper one is for $\epsilon_0=38.2$ GeV/fm$^3$
and the lower one is for $\epsilon_0=76.4$ GeV/fm$^3$.
}
\label{fig:lmbd1}
\end{figure}

To get a better understanding of the relation between
the observed pressure isotropization and radiative processes,
it is helpful to look at the fugacity evolution.
Define fugacity to be the ratio of particle density
to the equilibrium particle density, i.e., $\lambda=n/n_{eq}$.
When the temperature $T$ is taken to be the average
kinetic energy per degree of freedom, i.e., 
$T=\epsilon/(3 n)$, the resultant $n_{eq}$ leads
to the expression $\lambda=27\pi^2n^4/(16\epsilon^3)$.
For systems in thermodynamical equilibrium, i.e.,
in chemical and thermal equilibrium, 
$\lambda=1$.  The comparison of 
fugacity evolutions in Fig.~\ref{fig:lmbd1}
demonstrates the interplay between thermal equilibration
and chemical equilibration. 
Transverse initial conditions have evolutions lower
than their isotropic initial condition counterparts
mainly because of their slower early energy
density evolutions caused by the inside-outside
space-momentum correlation.
When only $2\rightarrow 2$ is
included, the system falls farther and farther away from
chemical equilibrium. As the $2\leftrightarrow 3$ processes
are turned on, the system is able to approach chemical
equilibrium. Even for transverse initial conditions
with $\epsilon_0=38.2$ GeV/fm$^3$ where there is
a decrease in fugacity mainly due to longitudinal
expansion, at late times, the fugacity
is able to reach a value that is comparable to other cases.
Note that in this case, there are more annihilations
than productions initially even when $\lambda\sim 1$ 
because of the transverse spatial distribution. This is
why the early evolution is slightly lower 
than the corresponding elastic only case.
It demonstrates that chemical equilibration depends
on kinetic equilibration in the radiative case.
For various cases with the $2\leftrightarrow 3$ processes,
the final fugacities at 2 fm/$c$ are all on the 80\% to 
90\% level. Note that the $\epsilon_0=38.2$ GeV/fm$^3$
case has an initial fugacity close to $1$, while the
$\epsilon_0=76.4$ GeV/fm$^3$ case is significantly 
undersaturated initially. As discussed before, this 
difference in initial fugacities is behind the observed
different energy dependences of pressure anisotropy evolutions
for the elastic only and the radiative cases.

\section{Summary and discussions}
\label{sec:summa}

The above study demonstrates the different behaviors of
evolutions with and without radiative processes. When
radiative processes are included, there is an approximate
$\alpha_s$ scaling that is in contrast with the 
$\epsilon_0 \alpha_s$ scaling seen when only elastic
processes are included. Evolutions with different initial
pressure anisotropies appear to approach the same
asymptotic evolution exponentially at late times.
If a partonic system is initially close to chemical 
equilibrium, radiative processes can significantly 
enhance pressure isotropization. For systems significantly 
undersaturated at the initial time, evolutions with
and without radiative processes can be close to each
other. This dependence on initial fugacity 
indicates that it may be necessary to 
use different $K$ factors
for elastic simulations at different collision energies.

\begin{figure}
\centering
\includegraphics{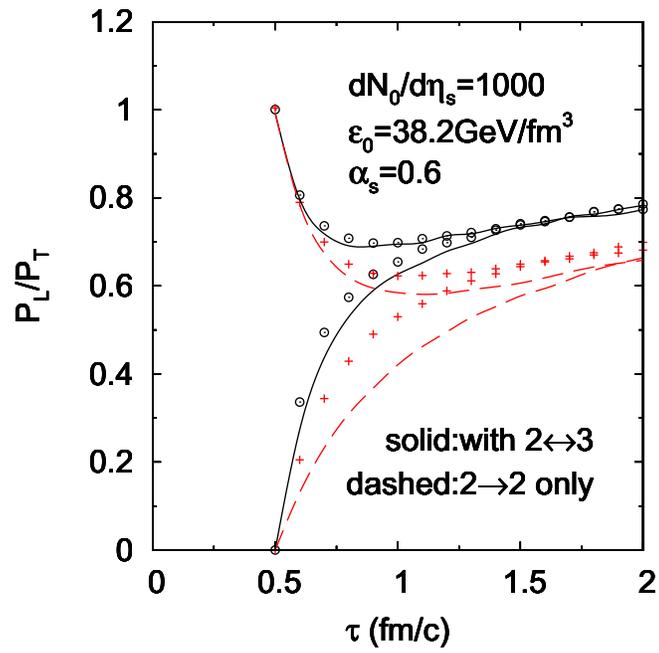}
\caption{Pressure anisotropy evolution for
$dN_0/d\eta_s=1000$,
$\epsilon_0=38.2$ GeV/fm$^3$, $\alpha_s=0.6$. Lines are
for exponential initial momentum distributions and
points are for condensate initial momentum distributions.
Solid lines and circles
are for the case including the $2\leftrightarrow 3$ processes
while dashed lines and pluses are for the case with 
$2\rightarrow 2$ only. 
}
\label{fig:plopt2}
\end{figure}

The above calculations start with initial exponential
momentum distributions. The dependence on the initial
momentum distribution can be studied by comparing the
results with those that start with initial momentum
distributions motivated by the Color Glass Condensate.
Fig.~\ref{fig:plopt2} has additional calculations
with step function initial momentum distributions.
A clear enhancement of pressure isotropization is
observed with the step function initial distributions
when only elastic processes are included. When
inelastic processes are included, there is not much
change in the pressure anisotropy evolution. In
other words, the pressure isotropization is robust 
against changes in the initial momentum distribution.
This difference between the elastic only
and with radiative reflects the faster thermalization
when radiative processes are included. Without
them, the step function momentum distribution
maintains its shape for a longer period of time,
resulting in a smaller screening mass and larger
collision rate per particle and a significant
change in the pressure anisotropy evolution.
This comparison demonstrates again
the importance of radiative processes in microscopic
simulations of relativistic heavy ion collisions.

\section*{Acknowledgments}
We thank S. Bass, R. Bellwied, F. Gelis, H. Grigoryan, M. Guenther, 
U. Heinz, S. Katz, V. Koch, M. Lisa, L. McLerran, H. Meyer, U. Mosel, 
S. Pratt, I. Vitev, H. Zhou
for helpful discussions and
the Parallel Distributed Systems Facilities of the
National Energy Research Scientific Computing Center
for providing computing resources.
This work was supported by the U.S. National Science Foundation
under grant No.'s PHY-0554930 and PHY-0970104.



%% References


%\bibliographystyle{elsarticle-num}



\end{document}